\documentclass[twocolumn]{aastex6}


\usepackage{graphicx}
\usepackage{dcolumn}
\usepackage{bm}
\usepackage{comment}
\begin{document}



\title{Illuminating Black Hole Binary Formation Channels with Spins in Advanced LIGO}

\author{Carl L. Rodriguez\altaffilmark{1,2}, Michael Zevin\altaffilmark{2},
Chris Pankow\altaffilmark{2}, Vasilliki Kalogera\altaffilmark{2}, and Frederic
A. Rasio\altaffilmark{2}}

\altaffiltext{1}{MIT-Kavli Institute for Astrophysics and Space Research, 77 Massachusetts
Avenue, 37-664H, Cambridge, MA 02139, USA}
\altaffiltext{2}{Center for Interdisciplinary Exploration and Research in Astrophysics (CIERA)
and Dept.~of Physics and Astronomy, Northwestern University
2145 Sheridan Rd, Evanston, IL 60208, USA}

\date{\today}

\begin{abstract}
The recent detections of the binary black hole mergers GW150914 and GW151226 have
inaugurated the field of gravitational-wave astronomy. For the two main
formation channels that have been proposed for these sources, isolated binary
evolution in galactic fields and dynamical formation in dense star clusters, the
predicted masses and merger rates overlap significantly, complicating any
astrophysical claims that rely on measured masses alone.  Here, we examine the distribution of spin-orbit
misalignments expected for binaries from the field and from dense star clusters.
Under standard assumptions for black-hole natal kicks, we find that black-hole
binaries similar to GW150914 could be formed with
significant spin-orbit misalignment only through dynamical processes.  In particular,
these heavy-black-hole binaries can only form with a significant spin-orbit \emph{anti}-alignment in the dynamical
channel.  Our results suggest that future detections of merging black hole
binaries with measurable spins will allow us to identify the main formation
channel for these systems. 
\end{abstract}

\maketitle


\section{Introduction}
\label{sec:level1}

The gravitational-wave detections GW150914 and GW151226 are the first direct
evidence of the formation and merger of
stellar-mass binary black holes (BBHs) in the local universe
\citep{Abbott2016a,Abbott2016b}.  Although many
channels have been explored for the formation of such systems, most proposals
fall into two categories: the ``field'' channel, in which BBHs are formed from
isolated stellar binaries, usually involving either a common-envelope phase
\cite[e.g.,][]{Voss2003,Dominik2012,Dominik2013,Belczynski2016} or chemically-homogeneous evolution due to rapid stellar
rotation \cite[e.g.][]{DeMink2016TheLIGO,Mandel2016MergingBinaries,Marchant2016},
or the ``dynamical'' channel, in which BBHs are created though
three-body encounters in dense star clusters
\citep[e.g.,][]{Sigurdsson1993,PortegiesZwart2000,Downing2010,Downing2011,Ziosi2014,Rodriguez2015a,Rodriguez2016a}.
Unfortunately, the masses and merger rates predicted by these models
often significantly overlap, making it difficult to discriminate between
different formation channels for BBHs even with multiple detections.  

However, masses and merger rates are not the only observable predictions from BBH
formation models.  In particular, the distribution of BH spin orientations are expected to
depend heavily on the binary formation mechanism.  For BBHs from the
field, it is expected that the individual BH spins should be mostly 
aligned with the orbital angular momentum \citep{Kalogera2000}, with any misalignment arising from
the momentum ``kick'' imparted to the orbit during core collapse.  For
dynamically-formed BBHs, both the spin and orbital
angular momenta should be randomly distributed on the sphere.  These
 spin-tilt misalignments produce relativistic
precession of the orbit, which can be detected through the amplitude modulations in the
gravitational waveform as the binary changes its orientation with respect to the detector \citep{Apostolatos1994,Vitale}.  

In this letter, we compare the expected distributions of spin-tilt misalignments
for binaries formed from isolated binary stellar evolution to
those formed from dynamical encounters in dense star clusters. We find that, for
sufficiently massive systems (such as GW150914), measurements of the BBH spin-tilt
 will allow LIGO to discriminate between dynamically- and field-formed
binaries.  In addition, we
find that dynamics provides the best route to forming binaries with a
significant component of the spins anti-aligned with the orbital
angular momentum.  Since Advanced LIGO can best constrain the component of the
spin angular momentum that is aligned with the orbital angular momentum
\cite[][and references therein]{TheLIGOScientificCollaboration2016} we suggest
that this may represent the best way to differentiate these BBH populations.

\section{Spin-Orbit Misalignment}

\begin{figure}[t!]
\centering
\includegraphics[scale=1.3, trim=0.in 0.3in 0in 0.0in,
clip=true]{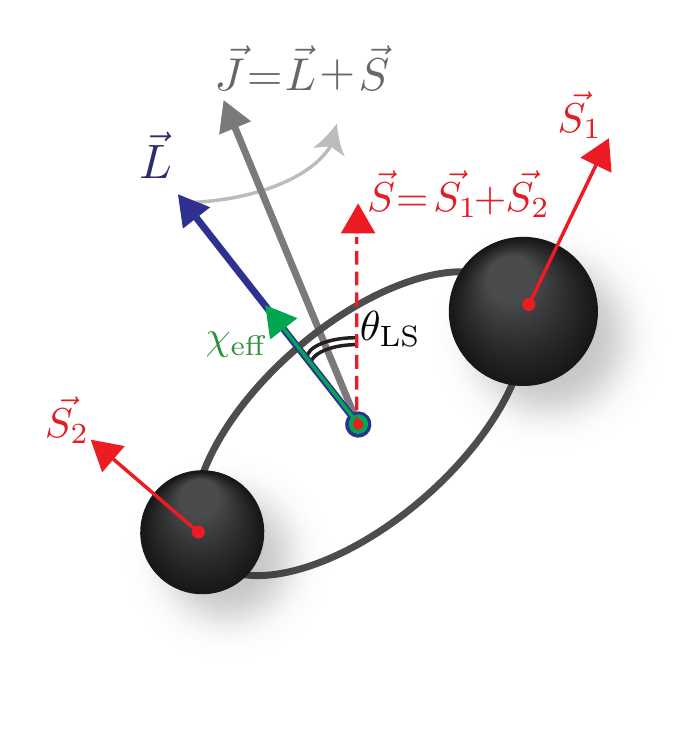}
\caption{Diagram of the vectors and angles that define the spinning BBH problem.
For any system where $\vec{S}$ and $\vec{L}$ are misaligned, the orbital plane will precess about the total angular momentum, $\vec{J}$.}
\label{fig:cartoon}
\end{figure}

In Figure \ref{fig:cartoon} we show the vectors and angles that describe 
the spin orientations of a BBH system.  For any binary in which
the total spin vector, $\vec{S}$, is misaligned with the orbital angular
momentum, $\vec{L}$, the
entire system will precess about the total angular momentum, $\vec{J}$
\citep{Apostolatos1994}.  We refer to the angle between $\hat{L}$ and $\hat{S}$,
$\theta_{\rm{LS}}$, as the spin-orbit misalignment, or the spin-tilt.  Although
restricting ourselves to only $\theta_{\rm{LS}}$ erases any information about the individual BH spins
and their potential resonant configurations \citep[such as those studied
in][]{Gerosaa,Gerosab,Trifiro}, it is the component of the mass-weighted spin angular
momentum perpendicular to the orbital plane, $\chi_{\rm{eff}} \equiv
\frac{c}{G(m_1+m_2)}\left[\frac{\vec{S}_1}{m_1}+\frac{\vec{S}_2}{m_2}\right]\cdot
\hat{L}$, that is best
constrained by Advanced LIGO.  The components of the
spins that lie in the plane of the orbit are 
responsible for the precession of $\hat{L}$ about $\hat{J}$, which induces
modulations in the amplitude of the gravitational waveform; however, the BBHs
detected by LIGO have so far not yielded significant constraints on the in-plane
spins of merging BHs \citep{Abbott2016e}.

\subsection{Misalignments in Isolated Field Binaries}

For BBHs formed from stellar binaries in the isolated channel, the primary
mechanism for inducing a misalignment between $\vec{L}$ and $\vec{S}$
 is the change in $\vec{L}$ imparted by the natal kick
 (NK).  The change in linear momentum instantaneously imparted to the
exploding star, from either emission of neutrinos or an assymetric explosion
mechanism \cite[e.g.,][]{Janka2013},  can significantly change the orbit of the
binary, resulting in a change to $\theta_{\rm{LS}}$.

We use the Binary Stellar Evolution \citep[BSE,][]{Hurley2002} code to create a BBH population representative of the field.  BSE uses a series of
metallicitiy-dependent fitted stellar tracks to rapidly model the evolution of stellar populations.  For binaries, BSE
also
models stable and unstable mass transfer, tidal
circularization, gravitational-wave emission, and the changes to the
orbital angular momentum arising from  kicks.  In addition, our version of BSE
contains several modifications to low-metallicitiy stellar winds
\citep{Vink2001,Belczynski2010} and core-collapse supernova (SN) \citep{Fryer2012} required to form ``heavy'' BBHs
such as GW150914 \citep{Abbott2016b}.

Our NK prescription is based on that developed in \cite{Fryer2012}.
Briefly, we assume that all compact objects receive a NK drawn a Maxwellian
distribution with a dispersion of
$\sigma = 265~\rm{km}/\rm{s}$ \citep[used to model the observed velocities of
pulsars in the galaxy,][]{Hobbs2005}.  However, for BHs we also assume that some
fraction of the mass ejected during core collapse will ``fall back'' onto the newly-formed proto-compact object.  When
this happens, conservation of momentum demands that the velocity of the BH be
reduced by: 

\begin{equation}
V_{\rm{kick}}^{\rm{BH}} = (1-f_{\rm{fallback}})V_{\rm{kick}}^{\rm{NS}}
\end{equation}

\noindent The fraction of material that falls back ($f_{\rm{fallback}}$) is
proportional to the core mass of the BH progenitor; for stars with large core
masses (those with a Carbon-Oxygen core mass greater than $11 M_{\odot}$), the rapid SN
prescription of \cite{Fryer2012} assumes complete fallback of material onto the
newly-formed BH (i.e.~$f_{\rm{fallback}}=1$).  This approximates the 
``direct
collapse'' of BH progenitors with proto-neutron star masses $\gtrsim 3M_{\odot}$
\citep{Fryer2001}.  A more-massive pre-collapse core
will produce a more-massive BH, accrete more fallback material, and experience a 
smaller NK.  We can expect that BBHs formed from isolated binaries
in the field should show less spin-orbit misalignment, especially as one
considers binaries with heavy BH components.

In addition, we consider two additional kick prescriptions for field BBH
populations.  Our proportional kick prescription assumes a much simpler relationship between
the new BH mass and the maximum NS mass, similar to that proposed for
neutrino-driven kicks \citep{Janka2013}:

\begin{equation}
V_{\rm{kick}}^{\rm{BH}} = \left(\frac{m_{\rm{NS}}}{m_{\rm{BH}}}\right)V_{\rm{kick}}^{\rm{NS}}
\end{equation}

\noindent where we assume $m_{\rm{NS}}=2.5M_{\odot}$.  We also consider the case where BHs recieve kicks comparable to those of NSs, i.e.:

\begin{equation}
V_{\rm{kick}}^{\rm{BH}} = V_{\rm{kick}}^{\rm{NS}}
\end{equation}

\noindent though as we discuss in Section \ref{sec:disc}, such kicks represent an extreme assumption not
supported by current observations.

In addition to varying the NK magnitudes, we also explore different NK directions.  By default, most population
synthesis studies have assumed kicks to be evenly distributed in solid angle
about the sphere of the exploding star.  However, pulsar observations have
suggested a correlation between the kick direction and the spin-axis of the
newly-formed proto-compact object \citep[e.g.,][]{Ng2007,Wang2006,Kaplan2008}.
Therefore, we consider two cases: an isotropic case, where NKs are distributed
randomly in solid angle over the sphere of the star, and a polar case, where we assume that all NKs are preferentially launched
from a cone with an opening angle of $10^{\circ}$ about the spin axis of each
star.   

For each kick prescription, we consider 11 different stellar metallicities: 1.5$Z_{\odot}$, $Z_{\odot}$, 0.5$Z_{\odot}$, 0.375$Z_{\odot}$, 0.25$Z_{\odot}$,
0.125$Z_{\odot}$,  0.05$Z_{\odot}$, 0.0375$Z_{\odot}$, 0.025$Z_{\odot}$,
0.0125$Z_{\odot}$, and 0.005$Z_{\odot}$.  We then evolve $10^5$ binaries in each
metallicitiy bin for 50 Myr with BSE.  We
sample the primary mass from
a $p(m)~dm \propto m^{-2.3}~dm$ power law from 18$M_{\odot}$ to $150M_{\odot} $\citep{Kroupa2001a}, and use a mass ratio distribution flat
from 0 to 1.  The eccentricities are drawn from a thermal distribution, $p(e)de
= 2e~de$, and the initial semi-major axes from a distribution flat in $\log(a)$ from $10R_{\odot}$
to $10^5R_{\odot}$.  We limit our sample to only those BBHs that will merge from
emission of gravitational waves in less than 13.8 Gyr.  

To assign spin-tilts to each binary, we assume that the only mechanism for
misaligning the orbital and spin angular momenta is the NK.  Although BSE
does not keep track of the three-dimensional spin-misalignments during its evolution, we record the angle
between the old and new $\hat{L}$ after each NK.  The total spin-tilt misalignment is
the combination of the two tilts.  We emphasize that this is a
highly-conservative estimate: both mass transfer during the common envelope
phase and tidal forces should
realign the spins of the first and second components of the binary between the
formation of the
first and second BHs.  However, this realignment would serve to decrease the large
spin-tilts reported here, making the distinction between field and cluster
populations even more distinct.  See Appendix \ref{app:A} for details.

Finally we evolve each
binary from a separation of $r=1000 (m_1+m_2) G/c^2$ to the separation where the
binary enters the LIGO band ($10~\rm{Hz}$) using the
precession-averaged post-Newtonian evolution in the
\texttt{Precession} package \citep{Gerosa2016} and assuming maximal spins for the BH components.  This was done to report the spin-tilt
misalignments that would be measurable by Advanced LIGO.  In practice, the 
spin-tilt distributions do not change between formation and $r=1000
(m_1+m_2) G/c^2$, since the couplings between $\hat{S}_1$, $\hat{S}_2$, and
$\hat{L}$ are $\mathcal{O}(v^6 / c^6)$ corrections to the spin evolution.
The differences in $\theta_{\rm{LS}}$ between $r=1000 (m_1+m_2) G/c^2$ and the
seperation each binary enters the LIGO band are also minor, but we report the
$10~\rm{Hz}$ values for comparison with previous results \citep{Gerosaa}.

\subsection{Cluster Binaries}
For BBHs formed in dense stellar environments, we assume the 
orbital and spin angular momenta are completely random.  As more
than 99\% of all BBHs that merge in the local universe from globular clusters are formed
through complicated and chaotic dynamical interactions \citep{Rodriguez2016b},
BBHs from clusters should have $\hat{L}$, $\hat{S_1}$, and $\hat{S_2}$
isotropically distributed across the sphere.  We conclude that
$p(\theta_{\rm{LS}})d\theta_{\rm{LS}} =
\sin(\theta_{\rm{LS}})/2~d\theta_{\rm{LS}}$, suggesting that clusters
preferentially form binaries with spins lying in the plane of the orbit.
Since
it has been shown \citep{Bogdanovic2007,Gerosa2015} that an isotropic
distribution of BBH spins remains isotropic during inspiral, we assume cluster
binaries to have randomly-distributed spins when they enter the LIGO band.  Note that we are only considering the ``classical'' channel of dynamical formation; many additional channels have  been proposed in which dynamics can induce mergers in BBHs that formed from pre-existing stellar binaries, such as those driven to merger via Kozai-Lidov oscillations
from either stellar-mass triples \cite[e.g.,][]{Silsbee2016} or binaries orbiting a super-massive BH \cite[e.g.,][]{Antonini2012,VanLandingham2016}, or from binaries that form and potentially accrete gas in
AGN disks \cite[e.g.,][]{Stone2016,Bartos2016}.  Although the spin distributions
of such scenarios are worthy of future study, for simplicity we do not consider them here.

\section{Results}

\begin{figure*}[tbh]
\centering
\includegraphics[scale=0.6, trim=0.9in 0.075in 0in 0.05in, clip=true]{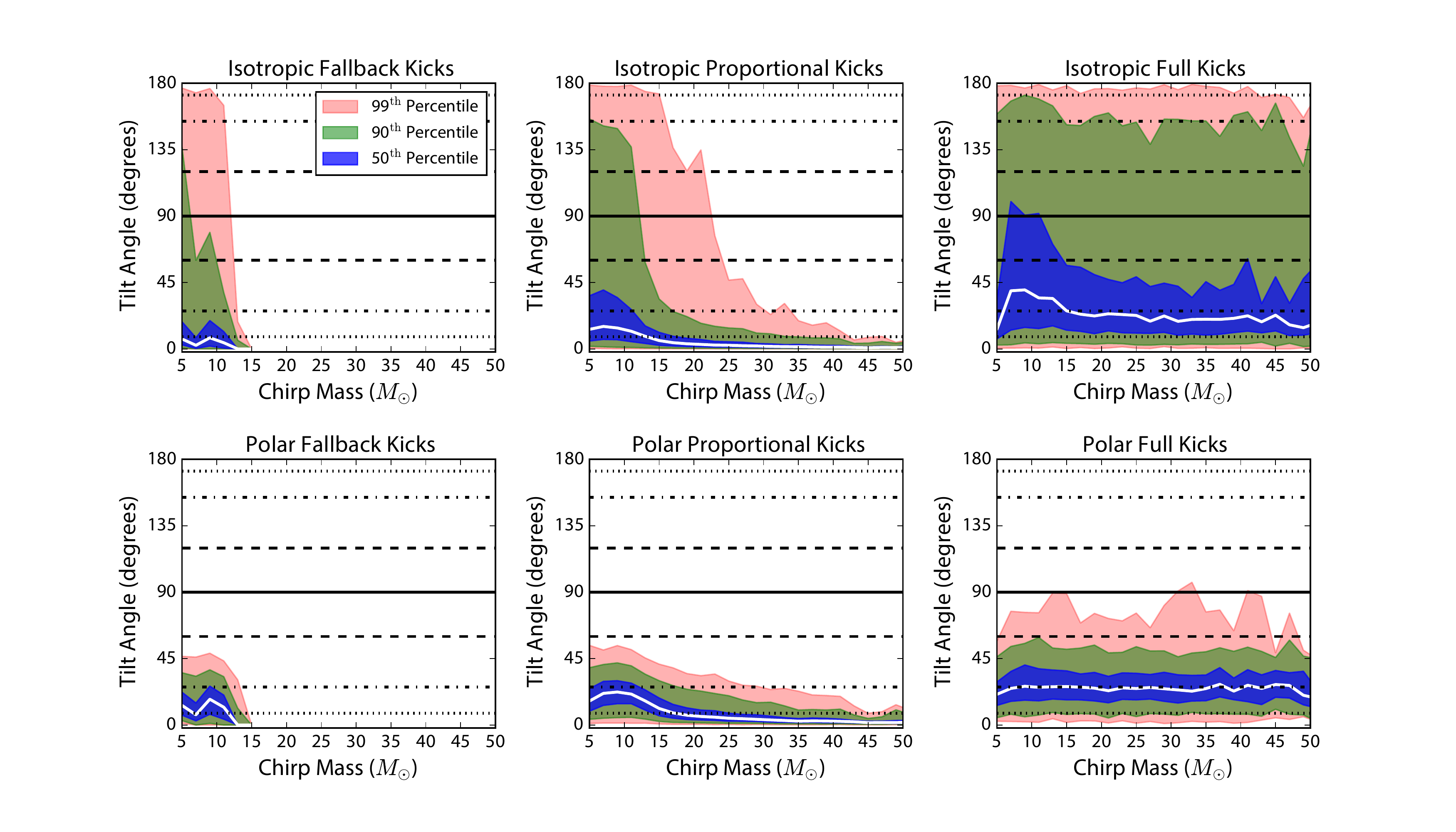}
\caption{The distribution of spin-tilt misalignments for our field and cluster
populations as a function of chirp mass.  The colors show the field population,
with the solid white line indicating the median value, and the blue, green, and
pink regions showing where 50, 90, and 99\% of sources lie in each $2M_{\odot}$
bin.  The distribution of cluster misalignments, evenly distributed in
$\sin{\theta_{\rm{LS}}}$, is shown in black, with the solid line indicating the
median, and the dashed, dot-dashed, and dotted lines showing the 50, 90, and
99\% regions respectively.  As we have explicitly assumed no realignment of the spins
between NKs, these represent the largest possible spin-tilts from the field
(see Figure \ref{fig:dist_2nd} for less conservative estimates).  Note that all binaries above $\sim
15M_{\odot}$ in the fallback prescription have zero spin-tilt misalignment, and
are not shown in the plot.}
\label{fig:dist}
\end{figure*}

Because the magnitude of the BH NK decreases with increasing BH mass, we
are most interested in the correlation between the spin-tilts and the binary
masses.  In Figure \ref{fig:dist}, we show the spin-tilt misalignment for each
of our six models of field binaries, overlaid with the (randomly distributed)
misalignments for BBHs from clusters.  As expected, both the fallback
and proportional prescriptions show a decreasing spin-tilt
misalignment as a function of binary chirp mass, defined as $\mathcal{M}_c
\equiv 
(m_1 m_2)^{3/5}/(m_1+m_2)^{1/5}$. Furthermore, for binaries that experience
polar kicks, the spin-orbit misalignments are limited to
$\theta_{\rm{LS}} \lesssim 90^{\circ}$ for the full NS kicks case, and
$\theta_{\rm{LS}} \lesssim 45^{\circ}$ for the fallback and proportional cases.
This behavior is to be expected: in order to anti-align the orbital and spin
angular momenta, the NKs must be able to reverse the orbital angular momentum,
which is best accomplished by a planar kick with sufficient speed to reverse the
orbital velocity.  However, the polar kick case explicitly excludes such
planar kicks.  The only exception would be the case where the first NK yields a
misalignment $\theta_{\rm{LS}}\sim 90^{\circ}$, placing the second star in a
position to emit a NK opposite to the direction of the orbital velocity.
However, such large first kicks frequently unbind the binary, and any BBHs that
survive are left with such large orbital separations that they will not merge
within a Hubble time.  The large tilts in the isotropic models are best
understood by decomposing the kick into two components: a polar component which
can torque the orbit up to 90◦, and a planar component, which can (in some
cases) reverse the orbital velocity, flipping the orbit by 180◦.  Because the
planar kick component can be launched in a direction opposite the orbital
velocity, the binary can be pushed into a tighter orbit, allowing it to merge
within a Hubble time.  On the other hand, the polar component of the kicks is
always tangential to the orbit (for the first kick), increasing the orbital
angular momentum and widening the orbit.  This creates a bias for small NKs and
correspondingly small tilts in the polar kick models, since only those systems
will remain bound and merge within a Hubble time \cite[as noted
by][]{Kalogera2000}.

Even when we allow for full-NS NKs independent of BH mass, the majority of
systems do not show tilts beyond $90^{\circ}$.  In Figure
\ref{fig:frac}, we show the fraction of BBHs in each model that have spin-tilts
greater than $90^{\circ}$ as a function of chirp mass.  For the polar kick
models, less than 1\% of binaries achieve a spin-orbit misalignment of greater
than $90^{\circ}$ at any given chirp mass.  For isotropic kicks, the possibility
of a spin-flip is significantly increased, since an isotropic distribution
allows for the planar kicks required to reverse the orbital velocity.  However,
these kick magnitudes must be on the order of and in the opposite direction to
the orbital velocity. For the isotropic fallback and isotropic proportional models, only 7\% and
10\% of the low-mass binaries ($\mathcal{M}_c \sim 5M_{\odot}$) have sufficient
kicks to flip the orbital angular momentum.  This fraction decreases as a
function of mass, such that the isotropic fallback model produces \emph{no} spin-orbit
misalignments for $\mathcal{M}_c \gtrsim 11M_{\odot}$, while $\sim$1\% of binaries
with $\mathcal{M}_c \sim 15M_{\odot}$ can yield $\theta_{\rm{LS}} > 90^{\circ}$.
Only the isotropic full-NS kick model can produce significant fractions
(10-30\%) of anti-aligned heavy BBHs.  For dynamically-formed binaries, 50\% of all systems show some anti-alignment of $\hat{S}$ and $\hat{L}$, as expected for systems whose angular momenta are isotropically distributed on the sphere.

\section{Discussion}
\label{sec:disc}

\begin{figure}[t]
\centering
\includegraphics[scale=0.7, trim=0in 0.05in 0in 0.05in, clip=true]{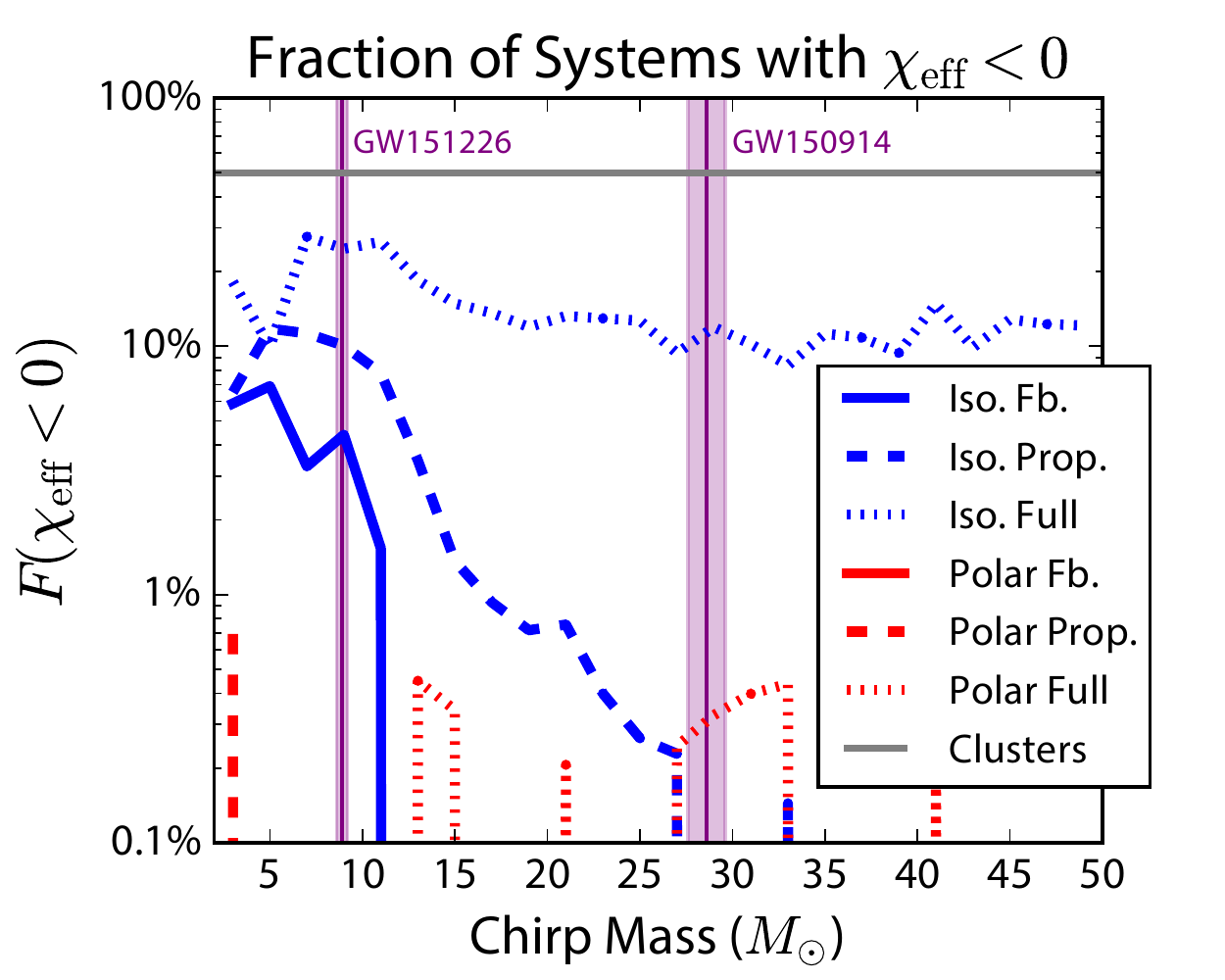}
\caption{Fraction of binaries from each field model and from clusters with
$\chi_{\rm{eff}} < 0$ as a function of chirp mass, with the median and 90\%
chirp masses of the two GW events in purple.  Since cluster spin-tilts are
distributed evenly in $\sin(\theta_{\rm{LS}})$, half of all systems will have
some component of the total spin anti-aligned with the orbital angular momentum.
For field populations, the fraction of systems with $\chi_{\rm{eff}} < 0$
decreases as a function of mass.. The only exception is the field model in which
BHs are given fully-isotropic kicks similar to neutron stars.  In that case,
10-30\% of sources can have spins partially anti-aligned with $\hat{L}$, regardless of mass.    For the polar kick models, only a
handful of binaries show misalignments greater than $90^{\circ}$.  This gives
rise to the small spikes at $\sim 0.5\%$ in the polar proportional and polar
full kick models.}
\label{fig:frac}
\end{figure}

Figure \ref{fig:frac} illustrates a key point of this letter: for sufficiently
massive binaries, the most efficient way to produce systems with spin components
anti-aligned with the orbital angular momentum is through dynamical encounters.
Parameter estimation of the lower-mass BBH detected by Advanced
LIGO, GW151226, suggests a chirp mass of $8.9^{+0.3}_{-0.3} M_{\odot}$ at the
90\% credible level \citep{Abbott2016e}, and shows significant evidence for BH spins that are
partially aligned with the orbital angular momentum.  Given the analysis here,
we cannot rule out either a field- or dynamical-formation scenario for GW151226.
On the other hand, GW150914, the most massive BBH merger detected to date, was
detected with a chirp mass of $28.1^{+1.8}_{-1.5}M_{\odot}$ at 90\% confidence
\citep{TheLIGOScientificCollaboration2016}.  
Our results suggest that, if a BBH similar to GW150914 were detected with a
measurably negative $\chi_{\rm{eff}}$, it would strongly suggest a dynamical
origin.  
Although parameter estimation of GW150914 hinted at such a configuration, with a
measured an aligned-spin value of $\chi_{\rm{eff}} = -0.09^{+0.19}_{-0.17}$ at a
90\% confidence, such a measurement does not definitively rule out either
large, in-plane spins (which would also arise from dynamical formation) or small,
aligned spins.

For systems similar to GW150914, Figure \ref{fig:dist} shows that only
full-NS NKs delivered in the plane of the orbit could produce a spin-orbit
misalignment greater than $90^{\circ}$.  However, we consider such kicks to be
highly unlikely.  Previous studies have indicated that
such strong NKs would reduce the BBH merger rate from dense stellar environments by an
order of magnitude  \citep{Rodriguez2016b}, and from the field by two orders of
magnitude \cite[][]{Dominik2013,Belczynski2016}.  Even under optimistic
assumptions, this would yield a combined merger rate of BBHs in the local
universe of $\sim 7~\rm{Gpc}^{-3}\rm{yr}^{-1}$, below the 90\% lower-limit of
$9~\rm{Gpc}^{-3}\rm{yr}^{-1}$ reported from the first observing run of Advanced
LIGO \citep{Abbott2016e}.  We conclude that it is unlikely that BHs can receive
such strong NKs across all mass ranges.  

It should be mentioned that recent analyses of low-mass X-ray binaries
(LMXBs) in the
galaxy have suggested that, while most BHs are consistent with no NKs at formation, at least a few BHs may receive NKs as high as
$\sim 100~\rm{km}/\rm{s}$ \citep{Podsiadlowski2002,Willems2005,Fragos2009,Wong2012a,Wong2014,Repetto2012,Repetto2015}.  In particular, \cite{Repetto2015} noted that
two LMXB systems, XTE J1118+48 and 0H1705-250, must have received kicks of at
least $\sim 100~\rm{km}/\rm{s}$ and $\sim 450~\rm{km}/\rm{s}$ respectively to
explain their current positions in the galaxy; however, all of the NK prescriptions employed
here can produce kicks of this magnitude for $5M_{\odot}-10M_{\odot}$ BHs (see Appendix
\ref{app:A}, Figure \ref{fig:speeds}), making our results consistent with the observed positions
of these LMXBs.  Furthermore, it has been shown that the estimated birth
velocity of H1705-250 can be explained by uncertainties in the observed position
of the LMXB, without the need to invoke such large NKs 
\citep{Mandel2016}.

Additionally, we have assumed that the amount of material that falls back on the
proto-compact object will reduce the velocity of the BH via conservation of
momentum.  However, it has been suggested that the fallback of material can
actually \emph{accelerate} the BH to speeds similar to neutron stars, either via
asymmetric accretion or though a gravitational ``tug-boat'' mechanism powered by
the asymmetric ejecta \citep{Janka2013}.  However, such behavior would still
only apply to BHs that eject some amount of material.  For heavy BBHs such as
GW150914, these prescriptions suggest that the BHs would form in a direct
collapse with no significant mass ejecta \citep{Fryer2001,Belczynski2016}.  A
direct collapse would also eliminate the possibility of an asymmetric SN
altering the spin of the compact object itself \cite[as has been invoked to explain the spin-misalignment of the double pulsar system PSR J0737−3039,][]{Farr2011a}.

\section{Conclusion}

In this letter, we explore the spin-tilt distributions of BBHs from different
formation channels.  We have shown for heavy BBH systems, such as GW150914, the
allowed range of spin-orbit misalignments that can be produced by BH NKs 
is limited.  Only under the extreme case where BHs of all masses can recieve NKs 
comparable to NSs, can isolated stellar evolution produce
spin-tilt misalignment greater than $90^{\circ}$.  On the other hand, BBHs
formed through dynamical processes in dense star clusters are
expected to produce isotropically-distributed spin-tilt misalignments, which
easily allow for the formation of BBHs with significantly anti-aligned spin and
orbital angular momenta.  Since any model of BH formation that allows for
full-NS NKs results in a predicted BBH merger rate below the 90\%
lower-limit observed by Advanced LIGO, we conclude that any sufficiently-massive
BBH merger ($\mathcal{M}_c \gtrsim 10-15M_{\odot}$, depending on the driving mechanism
of the NK) that shows a negative $\chi_{\rm{eff}}$ was most likely formed through dynamical processes.  

There are many additional facets of the BBH spin problem to be considered: first, although 50\%
of dynamically-formed binaries will have $\chi_{\rm{eff}}<0$ (assuming non-zero
component spins), this does \emph{not} mean that 50\% of binaries detected by
LIGO from clusters will have clearly discernible $\chi_{\rm{eff}}<0$.  Systems with
$\chi_{\rm{eff}}\gg 0$ are detectable at greater distances than systems with
$\chi_{\rm{eff}}\ll 0$ \citep{Ajith,Dominik2014}.   Furthermore, systems with
large spins in the plane of the orbit (the most
probable configuration for dynamically-formed binaries) will precess, producing 
amplitude modulations that can further decrease detectibility of
rapidly-spinning binaries.  Given that dynamics preferentially forms BBHs with
spins lying in the orbital plane, such precessional effects may offer the best
chance for identifying dynamically-formed BBHs.  Although precession has not
been observed in the two BBHs detected so far, improvements in the
lower-frequency limit of the LIGO instrument will increase the number of
precessional periods a binary experiences while in the LIGO band, significantly
improving the ability to measure the in-plane component of the BH spins.  Studies to fully characterize
the detection rate and distinguishability of these two astrophysical
populations \citep[similar to][]{Vitale2016,Stevenson2016} are currently underway.

\acknowledgments{
We thank Richard O'Shaughnessy, Chris Fryer, Will Farr, and Ilya Mandel for useful discussions, and
Davide Gerosa for making the \texttt{Precession} package public.  This work was
supported by NSF Grant AST-1312945, NSF Grant PHY-1307020,  and NASA Grant NNX14AP92G.  CR is grateful
for the hospitality of the Kavli Institute for Theoretical Physics, NSF Grant
PHY11-25915, and is supported by the MIT Pappalardo Fellowship in Physics. VK
and FAR also acknowledge support from NSF Grant PHY-1066293 at the Aspen Center
for Physics.}

\bibliographystyle{aasjournal}

\appendix
\section{Field Population}
\label{app:A}

\begin{figure*}[b]
\centering
\includegraphics[scale=0.6, trim=0.9in 0.075in 0in 0.05in, clip=true]{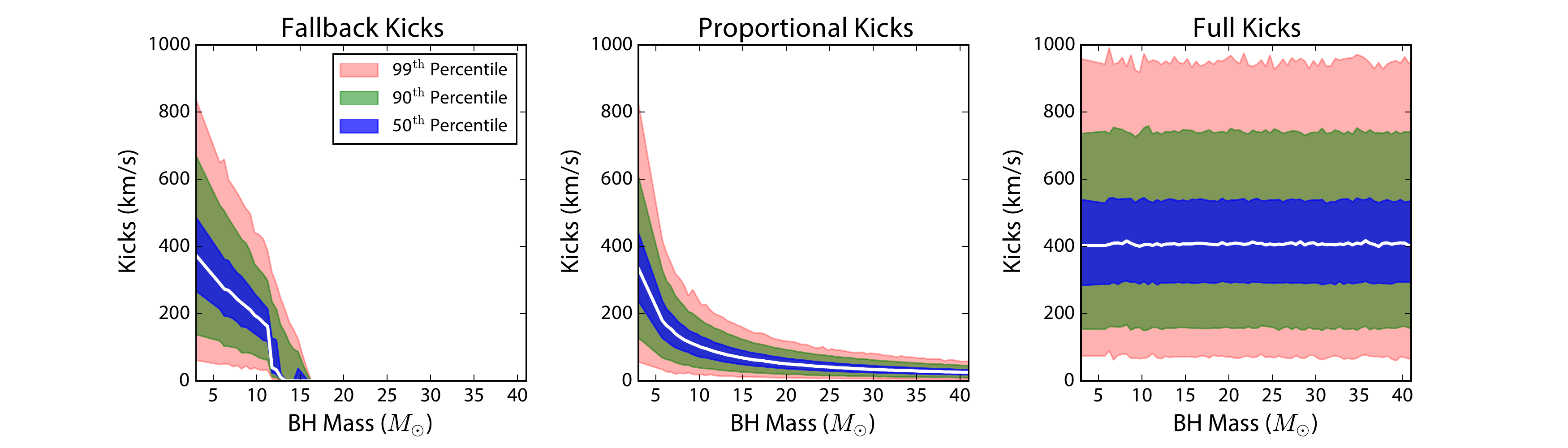}
\caption{Distribution of kick magnitudes for each of our three models as a
function of BH mass, for stars with initial masses from $20M_{\odot}$ to
$150M_{\odot}$ at a metallicitiy of $0.1Z_{\odot}$.  We include the median value
and percentile regions for each $0.5M_{\odot}$ bin.  The unusual behaivor of the
fallback prescription for BHs with masses between $11M_{\odot}$ and
$15M_{\odot}$ arises from the SN prescription developed in \citet[][Section
4]{Fryer2012}: BH progenitors with core masses from $6M_{\odot}$ to $7M_{\odot}$
experience full fallback of the SN ejecta, experiencing no NKs and producing BHs with masses in this range.  But BH progenitors with core masses from $7M_{\odot}$ to $11M_{\odot}$ eject some fraction of their mass, enabling non-zero kicks and decreasing the mass of the resultant BH.  This produces a bimodality in the fallback BH kicks.}
\label{fig:speeds}
\end{figure*}

Since BSE does not track the full three-dimensional orientations of the spin or
orbital angular momenta,
we record the tilt of the orbit after the formation of each BH.  This is done
using the formalism developed in the appendix of the original BSE paper
\citep{Hurley2002}.  Briefly, when BSE applies a kick to a binary, it assumes a
coordinate system with the non-exploding star at the origin and the exploding
star placed a position $r$ along the $\hat{y}$ axis.  The instantaneous orbital
velocity lies in the $x$-$y$ plane and the orbital angular momentum vector,
$\vec{L}$, points in the $\hat{z}$ direction.  When the SN occurs, a kick is
added to the orbital velocity, such that $\vec{V}_{\rm{new}} = \vec{V} +
\vec{V}_{\rm{kick}}$.  We show an example of the kick magnitudes for a single
stellar metallicity in Figure \ref{fig:speeds}.  Since we assume that the NK is applied instantaneously on the orbital timescale of the binary, the separation does not change.  The new direction of the orbital angular momentum vector is simply $\hat{L}_{\rm{new}} = \vec{r} \times \vec{V}_{\rm{new}} / |\vec{r} \times \vec{V}_{\rm{new}}|$, and the angle between the new and old angular momenta, $\nu$, is $\cos(\nu) = \hat{L}_{\rm{new}} \cdot \hat{z}$.

\begin{figure*}[htb]
\centering
\includegraphics[scale=0.6, trim=0.9in 0.075in 0in 0.05in, clip=true]{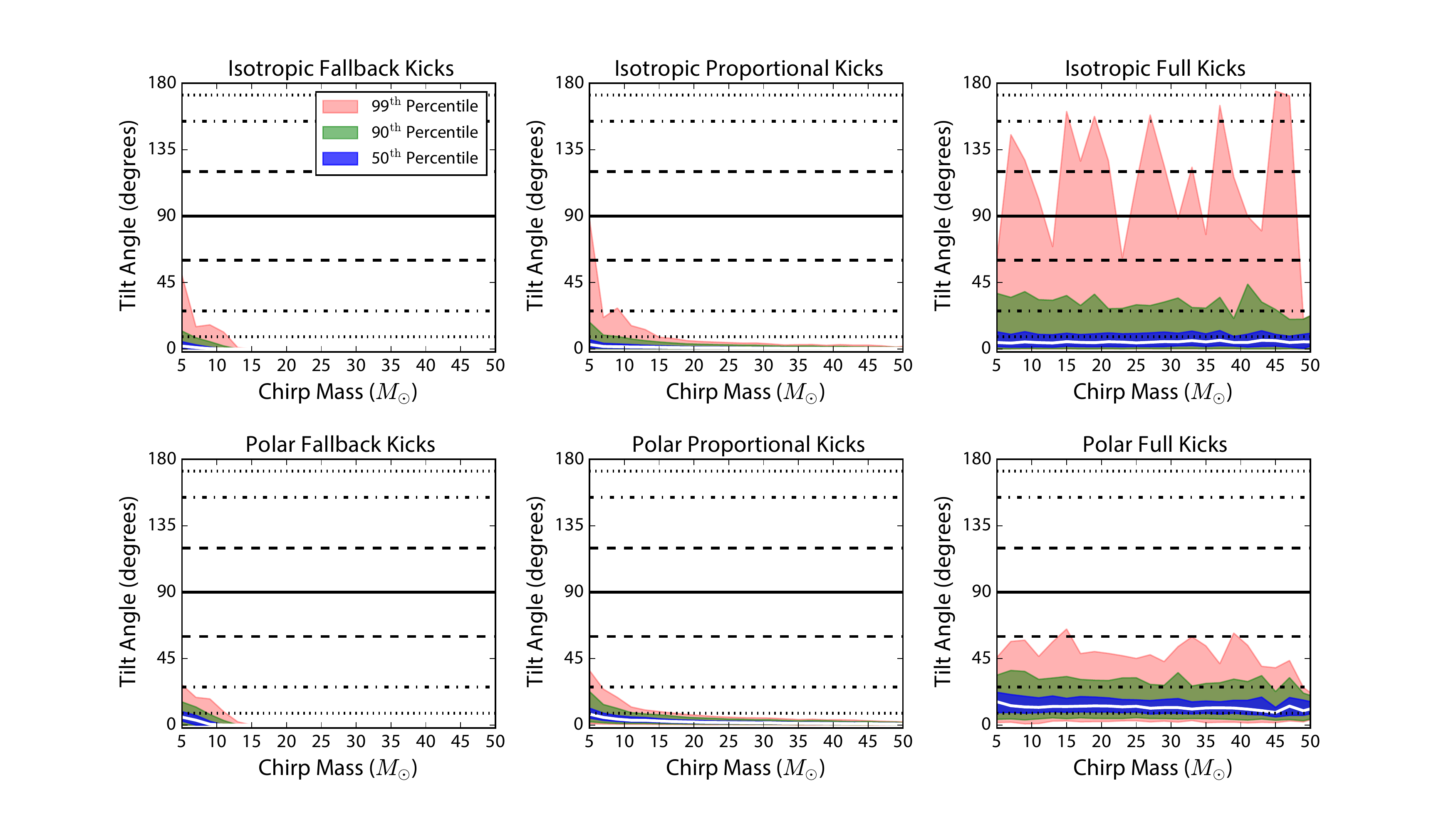}
\caption{Similar to Figure \ref{fig:dist}, but now assuming that mass transfer
and tidal torques can successfully realign $\vec{S}_1$ and $\vec{S}_2$ with
$\hat{L}$ between the first and second SNe.  Since the orbital velocities are
significantly larger prior to the second collapse than prior to the first
collapse, due to the decrease in separations following the common envelope
evolution, any NK for the second BH cannot change the orbit to the same degree
as the first NK, preventing the large spin-tilts observed in Figure \ref{fig:dist}.}
\label{fig:dist_2nd}
\end{figure*}

Because BSE considers the orbit-averaged evolution of the binary, we can track the angle that $\hat{L}_{\rm{new}}$ makes with $\hat{L}$, but not the phase of the projection of $\hat{L}$ into the orbital plane.  To that end, we select a random angle $\phi$ from $0$ to $2\pi$ for the orbital phase of $\hat{L}$ in the new orbital plane.  The total spin-misalignment is then

\begin{equation}
\cos{\theta_{\rm{LS}}} =  \cos(\nu_{1})\cos(\nu_{2}) + \sin(\nu_{1})\sin(\nu_{2})\cos(\phi)
\end{equation}

\noindent where $\nu_1$ and $\nu_2$ are the tilts created after the first and
second NKs.  Note that this is identical to equation 7 in \cite{Gerosaa}.  We
assume that the timescale between the formation of the first and second BHs is
sufficiently short that neither tides nor mass transfer can significantly
realign the spin of the second star between the two SN.  As such, our field
binaries all have $\hat{S_1}=\hat{S_2}$ by construction.  This is considered a
conservative assumption, as any physics which realigns $\vec{S}_{\rm{1}}$ or
$\vec{S}_{\rm{2}}$ with $\vec{L}$ will necessary produce smaller tilts  than
those reported in the main text.  As an illustration, we recreate Figure
\ref{fig:dist}, assuming that both spins realign with the orbital angular
momentum before the second BH forms (i.e.~$\theta_{\rm{LS}} = \nu_2$).  See Figure \ref{fig:dist_2nd}.

\end{document}